\documentclass{AGUJournal}
\draftfalse

\journalname{JGR-Planets}

\usepackage[utf8]{inputenc}
\usepackage[amsmath,upgreek]{}

\newcommand{\myadded}[1]{\ifdraft\textcolor{red}{#1}\else#1\fi}
\newcommand{\mydeleted}[1]{\ifdraft\textcolor{red}{\sout{#1}}\fi}
\newcommand{\myreplaced}[2]{\ifdraft\textcolor{red}{\sout{#1} #2}\else#2\fi}
\newcommand{\myexplain}[1]{\ifdraft\textcolor{red}{#1}\fi}

\begin{document}
	
	
	\title{Modeling the hydrological cycle in the atmosphere of Mars: Influence of a bimodal size distribution of aerosol nucleation particles}
	
	
	\authors{Dmitry S. Shaposhnikov\affil{1}, Alexander V. Rodin\affil{1,2}, Alexander S. Medvedev\affil{3}, Anna A. Fedorova\affil{2}, Takeshi Kuroda\affil{4,5} and Paul Hartogh\affil{3}}
	
	\affiliation{1}{Moscow Institute of Physics and Technology, Moscow, Russia}
	\affiliation{2}{Space Research Institute, Moscow, Russia}
	\affiliation{3}{Max Planck Institute for Solar System Research, G\"ottingen, Germany}
	\affiliation{4}{National Institute of Information and Communications Technology, Koganei, Japan}
	\affiliation{5}{Department of Geophysics, Tohoku University, Sendai, Japan}
	
	\correspondingauthor{D. S. Shaposhnikov}{shaposhnikov@phystech.edu}
	
	
	\begin{keypoints}
		\item A new microphysical scheme for water cycle on Mars implemented into a general circulation model.
		\item Accounting for bi-modality of aerosol particle distributions improves simulations of water ice \myreplaced{characteristics}{in the model} compared to observations.
		\item The fine fraction of atmospheric aerosols weakly affects simulations of water vapor in the model.
	\end{keypoints}
	
	
	\begin{abstract}
		
	We present a new implementation of the hydrological cycle scheme into a general circulation model of the Martian atmosphere. The model includes a semi-Lagrangian transport scheme for water vapor and ice, and accounts for microphysics of phase transitions between them. The hydrological scheme includes processes of saturation, nucleation, particle growth, sublimation and sedimentation under the assumption of a variable size distribution. The scheme has been implemented into the Max Planck Institute Martian general circulation model (MPI--MGCM) and tested assuming mono- and bimodal log-normal distributions of ice condensation nuclei. We present a comparison of the simulated annual variations, horizontal and vertical distributions of water vapor and ice clouds with the available observations from instruments onboard Mars orbiters. The accounting for bi-modality of aerosol particle distribution improves the simulations of the annual hydrological cycle, including predicted ice clouds mass, opacity, number density, particle radii. The increased number density and lower nucleation rates brings the simulated cloud opacities closer to observations. Simulations show a weak effect of the excess of small aerosol particles on the simulated water vapor distributions.
 		
	\end{abstract}
	
	
	\section{Introduction}
	
	Water in its different phases is a very important element of the current Martian climate, being a sensitive marker of meteorology in the atmosphere. It affects the Martian climate mostly through radiative effects of water ice clouds and scavenging dust from the atmosphere. Water was first detected in the Martian atmosphere more than a half century ago \citep{spinrad1963letter}. The next generation of studies broadly utilized data from orbiting and landing spacecraft, e.g., from Mars Atmospheric Water Detector (MAWD) onboard the Viking Orbiter \citep{jakosky1982seasonal}. To date, the main sources of information about the water distribution in the Martian atmosphere are the Thermal Emission Spectrometer (TES) onboard Mars Global Surveyor (MGS) \citep{smith2001thermal, smith2004interannual}, the Mars Climate Sounder (MCS) and the Compact Reconnaissance Imaging Spectrometer for Mars (CRISM) onboard Mars Reconnaissance Orbiter (MRO) \citep{smith2009compact}, the LIDAR instrument onboard the Phoenix Lander \citep{whiteway2009mars} and the Planetary Fourier Spectrometer (PFS), the Visible and Infrared Mineralogical Mapping Spectrometer (OMEGA) and the Spectroscopy for Investigation of Characteristics of the Atmosphere of Mars (SPICAM) instruments onboard Mars Express \citep{fedorova2006mars, fouchet2007martian, melchiorri2007water, tschimmel2008investigation, sindoni2011observations, maltagliati2011annual, trokhimovskiy2015mars}.
	
	The history of the water cycle modeling starts from the work of \citet{davies1981mars} \added{who developed the model to test the hypothesis that the observed seasonal and latitudinal distribution of water on Mars is controlled by the sublimation and condensation of surface ice deposits in the polar regions, and the meridional transport of water vapor. Then,} \citet{james1990role} \deleted{who }used a 1D model to show the role of water ice clouds in the water migration from north to south. The first comprehensive microphysical model of clouds was developed by \citet{michelangeli1993numerical} following the earlier attempts undertaken after measurements of water vapor vertical profiles \citep{kulikov1984modeling}. \citet{colaprete1999cloud} used microphysical models and \citet{haberle1999general} employed a Martian general circulation model (MGCM) to reproduce observations provided by Mars Pathfinder. \citet{richardson2002investigation} and \citet{richardson2002water} used the Geophysical Fluid Dynamics Laboratory (GFDL) MGCM to simulate the annual water cycle on Mars and compared it with the Viking MAWD data. Although the simulated climate was overly wet, these studies revealed the key mechanisms of the water transport. A more sophisticated model, which included transport, phase transitions and microphysical processes, has been developed by \citet{montmessin2004origin}. Later a microphysical model for Mars dust and ice clouds has been applied in combination with a model of the planetary boundary layer (PBL) for interpretation of measurements by the LIDAR instrument on the Phoenix Mars lander \citep{daerden2010simulating}. Observations of temperature inversions in the atmosphere of Mars \citep{hinson2004temperature} have motivated modelers to include effects of radiatively active water ice clouds (RAC) in MGCMs \citep{wilson2007diurnal, wilson2008influence, haberle2011radiative, madeleine2012influence, pottier2017unraveling}. These studies have demonstrated that accounting for RAC helped to reduce global temperature biases between simulations and observations at northern spring and summer \citep[see also the work of][]{urata2013simulations}. More MGCMs that include water cycle have been developed to date: DRAMATIC (Dynamics, RAdiation, MAterial Transport and their mutual InteraCtions) MGCM \citep{kuroda2017simulation}, NASA Ames GCM \citep{kahre2017updates}, GEM-Mars (The Global Environmental Multiscale model for Mars) GCM \citep{neary2018gem}, the Laboratoire de M\'et\'eorologie Dynamique (LMD) MGCM \citep{montmessin2004origin, navarro2014global} and the Oxford University MGCM \citep{steele2014seasonal}. The cloud scheme described by \citet{montmessin2004origin} was implemented at least in the latter two models, while the Oxford MGCM also used data assimilation scheme to nudge the simulated temperature to available observations.
	
	To successfully reproduce water cloud formations in the atmosphere of Mars, microphysical models require a correct \replaced{parameterization}{prediction} of the size distribution of aerosol particles, which serve as cloud condensation nuclei (CCN). Several observations have provided the evidence that this distribution is bimodal \citep{montmessin2006stellar, maattanen2013complete, fedorova2014evidence}. The distribution is called bimodal if its density function has two peaks, or modes. \citet{montmessin2002new} implemented such distribution into their one-dimensional model, prescribing two peaks with constant effective radii and variance for fine and large modes with a fixed ratio between them. They indicated that this assumption improved the simulations. For instance, it resulted in decrease of the effective radii of ice particles condensing on the CCN. In this study, we focus on the effects of the bimodal dust distribution on the global hydrological cycle.
	
	We present a new implementation of the hydrological scheme based on the \citet{montmessin2002new, montmessin2004origin} and \citet{navarro2014global} approaches into the Max Planck Institute (MPI) MGCM (also known as MAOAM -- Martian Atmosphere Observation and Modeling). This model with the employed physical parameterizations have been described in detail in the works of \citet{hartogh2005description, hartogh2007middle} and \citet{medvedev2007winter}. The most recent applications of this MGCM along with the current setup are presented in the works of \citet{medvedev2013dust, medvedev2015cooling, medvedev2016comparison} and \citet{yiugit2015gravity}. The earlier version of the hydrological scheme has been described in the work of \citep{shaposhnikov2016water}.
	
	The structure of this paper is as follows. In section~\ref{sec:MGCM}, we outline the newly developed hydrological scheme, discuss its main parameters and their importance. Section~\ref{sec:numerical} presents the simulations with the bimodal dust distribution and comparison with observations. In section~\ref{sec:mono}, we use different scenarios (with mono-modal and bimodal dust particle distribution) to demonstrate the sensitivity of the model to the bi-modality. Conclusions are summarized in section~\ref{sec:conclusion}.
	
	
	\section{Martian General Circulation Model}
	\label{sec:MGCM}
	
	
	\subsection{Transport}
	
	The MPI-MGCM employs a spectral dynamical core to solve the primitive equations of hydrodynamics on a sphere. The physics and tendencies are calculated on a 3D grid, and then are transformed into spectral coefficients at every time step ($\Delta t=$ 20 seconds). In the vertical, the grid is defined in the hybrid $\eta$-coordinate \citep{simmons1981energy} discretized into 50 levels, terrain-following near the surface and pressure based near the top, which was located at $\sim$100 km in our simulations. The horizontal grid is based on the Gauss-Kruger map projection with 32 and 64 bins in latitude and longitude, respectively. This discretization corresponds to a T21 triangular spectral truncation, which is a typical resolution of currently employed MGCMs, with a few exceptions for high-resolution experiments \citep{kuroda2015global, kuroda2016global, pottier2017unraveling}. Finite spatial resolution can be a source of numerically-induced features in simulations, which we discuss in the text.
	
	The spectral dynamical core is not well suited for simulation of the tracer transport. Instead, we adopted the advection based on a semi-Lagrangian explicit monotone second-order hybrid scheme and on the time splitting method in three spatial directions \citep{mingalev2010generalization}. A thorough examination of performed runs has confirmed that this scheme maintains a high order of conservation of water masses and solution accuracy appropriate for general circulation modeling \citep{shaposhnikov2016water}. In addition to advection, transport includes diffusion and mixing associated with subgrid-scale processes. The importance of vertical eddy mixing for modeling the water cycle was emphasized by \citet{richardson2002investigation}. In our simulations, we use the Crank-Nicolson implicit method with the Richardson number-based diffusion coefficients \citep{becker2007nonlinear} to solve the vertical diffusion equation.
	
	
	\subsection{Bimodal dust distribution}
	\label{sec:bimod-dust}
	
	We employ a predetermined dust scenario that represents a seasonal evolution of the zonally averaged aerosol optical depth $\tau$ in the thermal IR based on MGS-TES and MEX-PFS measurements with the global dust storms removed (the so-called ``MGCM dust scenario'' \citep{medvedev2011influence}). Thus, dust is not transported in the model. We use four bins to represent the cloud condensation nuclei size distribution: 0.03, 0.1, 0.3, 1 $\mathrm{\mu}$m. For the ice particle size distribution, a two-moment scheme is used at every bin, for keeping separately track of the ice mass and number of particles \citep{rodin2002moment}. Note that the particles were assumed to be spheres, although it may be a poor approximation for smallest nuclei. The size of ice particles determines some of their other microphysical properties and the sedimentation rate. The latter is accounted for through corrections to the advective vertical velocity for every particle bin according to the formula obtained by \citet{korablev1993vertical} with the Cunningham correction. Then, the CCN number density in each bin can be calculated from, for instance, the bimodal log-normal dust distribution \citep{fedorova2014evidence}, as shown in Figure~\ref{fig:bimodal_scheme}.
	
	\begin{figure}[h]
		\setcounter{figure}{0}
		\centerline{\includegraphics[width=30pc]{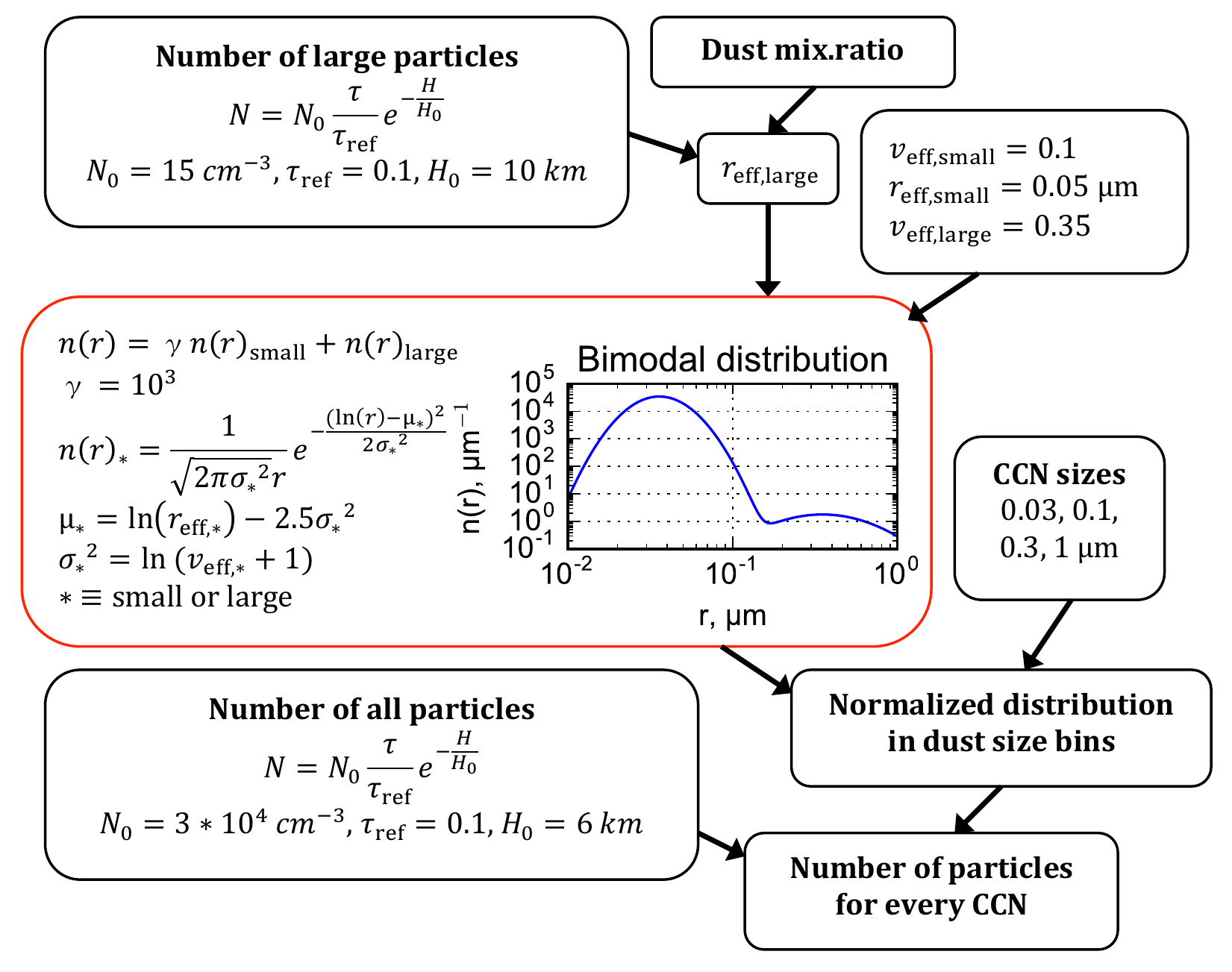}}
		\caption{The operation scheme for calculating the bimodal dust distribution. $N$ is the number density of large/small particles, $\tau$ is the dust optical depth in a vertical column (``MGCM dust scenario''), $H$ is the altitude of the grid cell, $r_\textrm{eff}$ is the effective radius, $v_\textrm{eff}$ is the effective variance, $\gamma$ is the population ratio, $n(r)$ and $n(r)_*$ are the bimodal and mono-modal size distribution functions, correspondingly.}
		\label{fig:bimodal_scheme}
	\end{figure}
	
	Following \citet{montmessin2002new}, the mono-modal log-normal density function $n(r)_*$ can be defined as
	\begin{linenomath*}
		\begin{equation}
		n(r)_* = \frac{1}{\sqrt{2 \pi \sigma_*^2} r} \exp{\left(-\frac{(\ln{r}-\mu_*)^2}{2 \sigma_*^2}\right)},
		\end{equation}
	\end{linenomath*}
	where the index $_*$ indicates the fine or large modes, respectively, and the parameters $\sigma_*$ and $\mu_*$ are related to the effective radius $r_\textrm{eff,*}$ and variance $v_\textrm{eff,*}$:
	\begin{linenomath*}
		\begin{equation}
		r_\textrm{eff,*} = \exp{\left( \mu_* + \frac{5}{2} \sigma_*^2 \right)}, \quad
		v_\textrm{eff,*} = \exp{\left( \sigma_*^2 \right)} - 1.
		\end{equation}
	\end{linenomath*}
	
	Then, the procedure of number density calculation for each bin includes the following steps:
	\begin{enumerate}
		\item Number density of large particles and the dust mass mixing ratio in the cell vary according to the annual dust scenario. Following the work of \citet{montmessin2004origin}, the number density $N_\textrm{large}$ of large particles can be calculated from the equation
		\begin{linenomath*}
			\begin{equation}
			\label{eq:num_density}
			N_* = N_0 \frac{\tau}{\tau_\textrm{ref}} e^{-\frac{H}{H_0}},
			\end{equation}
		\end{linenomath*}
		where $H$ is the altitude of the cell; $N_0$ = 15 cm$^{-3}$, $\tau_\textrm{ref}$ = 0.1, $H_0$ = 10 km are the adopted in this work parameters. The dust mass mixing ratio $Q$ in the cell is produced by the ``MGCM dust scenario'' and \replaced{vary}{varies} in time and space.
		
		\item The effective radius for large particles, $r_\textrm{eff,large}$, is calculated following \citep{maltagliati2011annual}:
		\begin{linenomath*}
			\begin{equation}
			r_\textrm{eff,*} = \left( \frac{Q}{4/3 \pi \rho_\textrm{ice} \rho_\textrm{air} N_*} \right)^\frac{1}{3} \left( 1 + v_\textrm{eff,*} \right),
			\end{equation}
		\end{linenomath*}
		where $v_\textrm{eff,large}$ = 0.35, $\rho_\textrm{ice}$ and $\rho_\textrm{air}$ are values of density of ice and air, respectively.
		
		\item Using constant values for the effective variance and radius of small particles $r_\textrm{eff,small}$ = 0.05 $\mathrm{\mu}$m and $v_\textrm{eff,small}$ = 0.1, the probability density function of the bimodal distribution $n(r)$ is obtained as a sum of $\gamma*n(r)_\textrm{small}$ and $n(r)_\textrm{large}$. Although \added{observations indicate that} the population ratio $\gamma$ may vary in time and space (see Figure~\ref{fig:maoam_bimodal_evolution}b), we adopt the constant $\gamma = 10^3$ in the bimodal case for all altitudes. The importance of this assumption can be a subject of future studies.
		
		\item Having the CCN sizes and number density of all particles, it is straightforward to calculate the number of particles in each bin and use it in the microphysical equations. For that, one has to specify the number density of small particles $N_\textrm{small}$ (which will be close to the number density of all particles in the case of large population ratio) with the equation~(\ref{eq:num_density}) and parameters $N_0$ = 3$\times$10$^4$ cm$^{-3}$, $\tau_\textrm{ref}$ = 0.1, $H_0$ = 6 km. All the mentioned parameters were selected according to the work of \citet{fedorova2014evidence}.
		
	\end{enumerate}
	
	\begin{figure}[h]
		\setcounter{figure}{1}
		\centerline{\includegraphics[width=30pc]{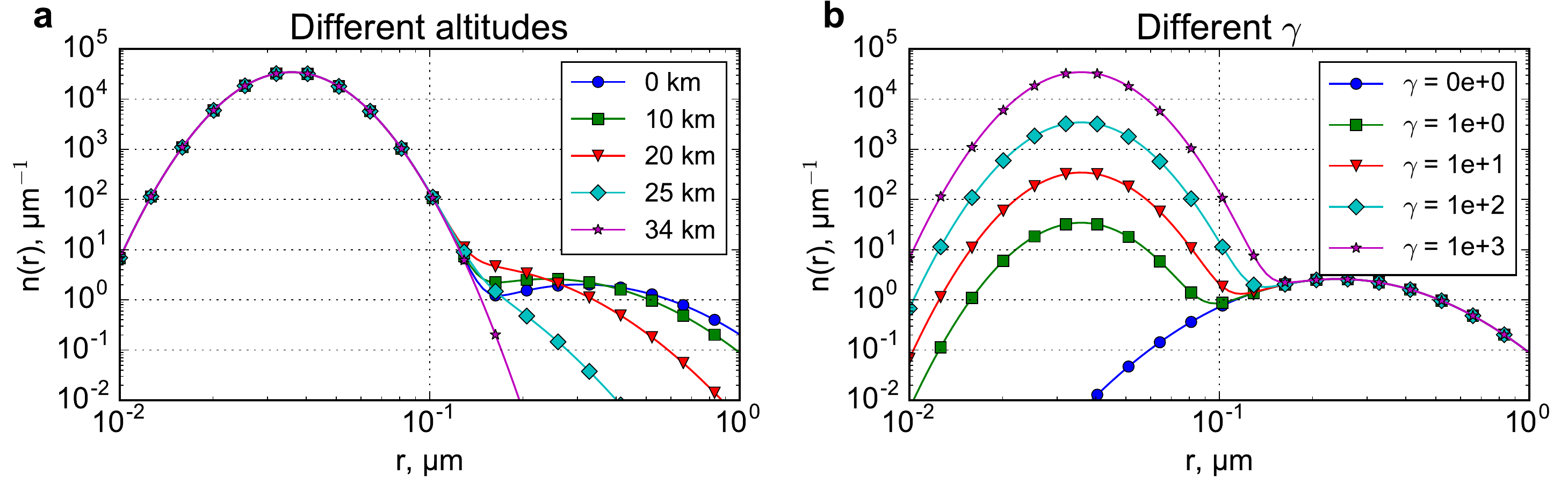}}
		\caption{Evolution of the probability density function of the bimodal aerosol particles distribution with a) altitude and b) population ratio $\gamma$ in the MPI-MGCM averaged over latitude and longitude around $L_s=120^{\circ}$. Different curves correspond to a) different altitudes and b) different $\gamma$ (see the legend).}
		\label{fig:maoam_bimodal_evolution}
	\end{figure}
	
	We use a bimodal distribution that includes much smaller particles than suggested by \citet{montmessin2002new}. The effective radius of the fine mode is only 0.03 $\mathrm{\mu}$m and the number density ratio up to 10$^3$-10$^4$ compared to $\sim$0.2 $\mathrm{\mu}$m and 25, correspondingly. As it was described above, the assumed distribution has a variable effective radius for the large mode in the \replaced{addopted}{adopted} fixed dust scenario. This allows for taking into account a variation of the distribution of large particles and their size decrease with height. Figure~\ref{fig:maoam_bimodal_evolution}a demonstrates that the second peak corresponding to the large mode shifts to the left with increasing altitude. As a result, above 40--60 km (depending on latitude and season), two modes merge and the distribution appears as a mono-modal, with small particles only. Some studies suggested that the source of such particles can be a micrometeoric smoke \citep{bardeen2008numerical}, \replaced{and}{while others assumed} that the corresponding particle lifting can be caused by electrical forces \citep{kok2006enhancement}. Other studies indicate that eddy diffusion and dynamic effects are more plausible sources of the fine mode \citep{fedorova2014evidence}.
	
	Note that to date, the dust bi-modality was reliably observed only in some seasons \citep{montmessin2006stellar, maattanen2013complete, fedorova2014evidence}. Therefore, the assumption of bi-modality throughout the entire Martian year still must be validated with measurements. The simulations presented here demonstrate the sensitivity of the water cycle in the model to \replaced{this input parameter}{bi-modality}, and the noticeable improvements it produces.
	
	
	\subsection{Nucleation}
	
	The water vapor saturation pressure in a given spatial grid cell provides information about the amount of the gaseous water held by the air parcel. In the model, the saturation pressure $p_\textrm{sat}$ is expressed as a function of temperature $T$ with a modification of the August-Roche-Magnus formula \citep{curry1998thermodynamics}:
	\begin{linenomath*}
		\begin{equation}
		p_\textrm{sat} = 611\;\textrm{Pa}\;\exp\left(22.5 \left(1-\frac{273.16\;\textrm{K}}{T}\right)\right).
		\end{equation}
	\end{linenomath*}
	
	Ice clouds are formed whenever water vapor nucleates on CCN. For quantification of water ice condensation on dust nuclei, a heterogeneous nucleation rate is evaluated according to \citet{jacobson2005fundamentals}:
	\begin{linenomath*}
		\begin{equation}
		J_\textrm{het} = 4 \pi r_c^2 \beta^2 Z_\textrm{n(het)} \Delta t \exp{\left(\frac{-\Delta G^{*}_\textrm{het}}{k_b T}\right)},
		\label{eq:J_het}
		\end{equation}
	\end{linenomath*}
	where $J_\textrm{het}$ is the heterogeneous nucleation rate (number of embryos, cm$^{-2}$~s$^{-1}$) of new particles on a surface, $r_c$ is the critical radius, $\beta$ is the number of gas molecules striking the substrate surface per second, $\Delta t$ = 10$^{-13}$ s is the characteristic time that a gas molecule spends on the surface before bouncing off, $Z_\textrm{n(het)}$ and $\Delta G^{*}_\textrm{het}$ are the Zeldovich nonequilibrium factor and the critical change in Gibbs free energy, respectively (adjusted for heterogeneity), $k_b$ is the Boltzmann constant, $T$ is the temperature.
	
	The number of active CCN also depends on the parameter $m_h = \cos{\theta}$, where $\theta$ is the ``contact angle'' (for the liquid phase, it is the angle of the interface between the droplet and its nucleus). We use the value $m_h = 0.96$ in calculations, which is consistent with the typical range between 0.93 and 0.97 for Martian dust \citep{trainer2008measurements}. There are indications that the value of this parameter significantly affects the distribution of water in simulations \citep{navarro2014global}. Although the laboratory studies indicate that the contact parameter is temperature dependent \citep{iraci2010water}, our numerical experiments with the constant contact angle produce results that better fit the observations compared to the previously used parameterizations like those described by \citet{maattanen2014estimating}. Since these parameterizations differ for particles of different type, shape and size, we select the simplest one with a fixed contact angle.
	
	
	\subsection{Particle Growth and Sublimation}
	
	Once water vapor is condensed, the ice particles are prone to growth determined by various factors. The rate of particle growth $\frac{dr}{dt}$ depends on the current water vapor saturation ratio $S$, the saturation pressure ratio over a curved surface $S_{eq}$, the molecular $F_D$ and the heat $F_H$ diffusion resistances \citep{montmessin2002new}:
	\begin{linenomath*}
		\begin{equation}
		\frac{dr}{dt} = \frac{S - S_{eq}}{r (F_D + F_H)}.
		\label{eq:drdt}
		\end{equation}
	\end{linenomath*}
	The resistances in (\ref{eq:drdt}) reflect the contribution of the molecular and the heat diffusion in the condensation process:
	\begin{linenomath*}
		\begin{equation}
		F_D = \frac{\rho_{ice} R T}{D'_i M_{H_2O} p_{sat}},
		\label{eq:f_d}
		\end{equation}
		\begin{equation}
		F_H = \frac{L_{ice} \rho_{ice}}{K_{air} T} \left(\frac{L_{ice} M_{H_2O}}{R T} - 1\right),
		\label{eq:f_h}
		\end{equation}
	\end{linenomath*}
	where $R$ is the universal gas constant, $M_{H_2O}$ is the water molar mass, $L_{ice}$ and $K_{air}$ are the temperature dependent latent heat of ice and thermal conductivity of air, respectively, parameterized in accordance with \citet{murphy2005review} and \citet{gori2004theoretical}. Here $D'_i$ is the molecular diffusion coefficient corrected for collision geometry and sticking probability \citep[Eqns. 16.17--16.19]{jacobson2005fundamentals}:
	\begin{linenomath*}
		\begin{equation}
		D'_i = D \left[1 + Kn_i \left( \frac{1.33 + 0.71 Kn_i^{-1}}{1 + Kn_i^{-1}} + \frac{4 (1 - \alpha_i)}{3 \alpha_i} \right) \right]^{-1},
		\label{eq:D'_i}
		\end{equation}
	\end{linenomath*}
	where $D$ is the molecular diffusion coefficient of water vapor in the Martian atmosphere, $Kn_i$ corresponds to the Knudsen number of the condensing vapor with respect to particles of size $i$ and $\alpha_i$ is the mass accommodation (sticking) coefficient of vapor obtained by \citet{kong2014water}.
	
	Ultimately, the hydrological cycle in the Martian atmosphere is critically dependent on the water sublimating from the surface. The model accounts for this with the turbulent flux at the bottom of the atmosphere $E_w$ \citep{montmessin2004origin}:
	\begin{linenomath*}
		\begin{equation}
		E_w = \rho_\textrm{air} C_d u_{*}(q_{vg}-q_{va}),
		\end{equation}
	\end{linenomath*}
	where the friction velocity $u_{*}$ depends on zonal and meridional wind velocities at the lower layer of the atmosphere, the variable $q_{vg}$ describes the water vapor partial pressure at this level, and $q_{va}$ is the saturation mixing ratio at the midpoint of the bottom layer. The saturation ratios are directly determined from the local pressure. The drag coefficient $C_d$ is set to $0.005$. In the model, the northern polar cap is the only continuous source of water in the atmosphere, specified as an infinite reservoir of the surface ice northward of 80$^{\circ}$N. This latitude corresponds to the middle of the model grid closest to the extent of the observed cap. The shape of the constant ice cap is a disc, while condensation and sublimation can modify temporarily the surface ice in other regions over the course of simulations. The MPI-MGCM uses the observed surface thermal inertia and albedo everywhere, except at the ice caps, where the albedo is set to 0.3 \citep{hartogh2005description, medvedev2007winter}.
	
	
	\section{Simulations}
	\label{sec:numerical}
	
	The model runs have been initialized with the distribution of water vapor in the atmosphere obtained in our earlier experiments \citep{shaposhnikov2016water}. As a reminder, the previous simulations started with a linear latitudinal gradient of water vapor (from 0 in the south to 200 ppm in the north) and no ice \added{outside the caps} was prescribed anywhere in the atmosphere at the beginning. Then, runs have been performed for several Martian years. The results to be presented here are based on daily averaged quantities, if not stated otherwise, and represent the second model year of simulations, counting from the restart. \replaced{In this paper we explore processes, whose timescales are much shorter than one Martian year and do not require long-term stability of the simulated hydrological cycle. The issues of seasonal cycle repeatability are beyond the scope of this paper.}{Water production over the shown year differs from that of the previous model year within a few percent and alternate signaling that the model seems to achieve a quasi-stable state. Details of the hydrological cycle repeatability and interannual variations are beyond the scope of the paper, here we only prove that the model meets general requirements of water mass conservation, numerical stability and accuracy.}
	
	
	\subsection{Annual Variations}
	\label{sec:annual}
	
	Figure~\ref{fig:compare_h2o_sum} presents a comparison of the simulated water vapor column density and the one observed with SPICAM. The SPICAM data are an average over 5 Martian years (MY27--31). The model reproduces both the seasonal asymmetry of the hydrological cycle and the total amount of water vapor in the atmosphere. The discrepancy between the observed and simulated quantities does not exceed $\sim$10 pr.$\mathrm{\mu}$m, and varies with seasons. The maximum amount of vapor occurs near the north pole between $L_s=90^\circ$ and 140$^{\circ}$. The model also reproduces the specific southward migration afterwards. The excess of simulated water vapor column at $L_s=170^\circ - 180^\circ$ in 0$^\circ$N -- 45$^\circ$N can be explained by an overly strong vertical eddy diffusion, which transports water vapor over the subliming polar cap. Another deficiency of the model results compared to SPICAM is the lack of vapor after $L_s \sim$270$^\circ$. Near the south pole, this can be related to the restriction imposed on the surface temperature. Southward of 85$^{\circ}$S, CO$_2$ ice is assumed to be present permanently on the surface, thus preventing temperature increase. It is also seen that the modeled cycle is somewhat delayed compared to the observations. The simulated scarcity of water in middle latitudes near $L_s \sim$300$^\circ$ could be explained either by inadequate meridional eddy transport, or insufficient evaporation of the south polar cap in the ``mean'' dust scenario lacking the seasonal dust storm. Despite the mentioned discrepancies, the simulated water cycle generally reproduces the observations well (the root mean square error is 4.3 pr.$\mathrm{\mu}$m). The simulated water vapor column density in the summer season at the north pole with the maximum near 50-70 pr.$\mathrm{\mu}$m agrees with the observations and other GCMs. The simulated vapor is also absent in the north and south polar regions during winter. The model exhibits a good repeatability of the vapor cycle, as can be judged by the match between the end and beginning of the model year.
	
	Further comparison of the simulated water vapor column with SPICAM data for the particular Martian years is shown in Figure~\ref{fig:maoam_h2o_lat_compar}. The upper panel displays the comparison at 70$^{\circ}$N, where the amount of water vapor is strongly affected by sublimation from the north polar cap. It illustrates the time lag of the simulated vapor in the atmosphere with respect to the observations. The main reason of this is presumably the delayed ice cap sublimation caused by low surface temperatures. Nevertheless, the simulated total vapor column is in a good agreement within the observed year-to-year variability.
	
	The middle panel shows a reasonably good agreement of the simulations and observations at the equator, with observed interannual variations being much smaller than in midlatitudes. The best agreement between the MGCM and observations occurs in MY28 and the worst in MY29. Regarding the latter, it is important to mention that the observed values were most scattered also during the observational period in MY29. Note that, near $L_s=100^{\circ}$, observations for MY27 and MY30 show a rapid growth of water vapor up to 15 pr.$\mathrm{\mu}$m, which is not reproduced in the model, and for which there are no observational data in other years. This may be linked to the transient baroclinic wave activity \citep{barnes1993mars, kuroda2007seasonal}.
	
	The lower panel corresponds to 55$^{\circ}$S, i.e., to the region influenced by the south ice polar cap. Figure~\ref{fig:maoam_h2o_lat_compar}c shows some lack of water vapor in the simulations during $L_s=270^{\circ} - 360^{\circ}$. However, this simulation is very similar to the observations in MY28, when the Martian atmosphere was affected by the global dust storm. As it was mentioned in section~\ref{sec:bimod-dust}, we do not include dust storms in the MGCM dust scenario, and the apparent agreement between the model and observations during $L_s=270^{\circ} - 360^{\circ}$ on 55$^{\circ}$S is accidental. The lack of vapor at this latitude can indicate weaker sublimation from the south pole cap. Finally, the regular spikes in the simulated vapor between $L_s=220^{\circ}$ and 260$^{\circ}$ are caused by the finite size of model grid cells. This phenomenon is the result of a staged disappearance of seasonal frost, which is imposed by the sparse model gridding. Sublimation can be fast enough to completely remove the surface frost from a grid cell before its southward neighbor is exposed to sunlight. Other models with a comparable spatial resolution demonstrate similar effects (see, for example, Figure~1b of \citet{lefevre2008heterogeneous}).
	
	Because SPICAM measurements of water ice are not as plentiful as of vapor, we compare the simulated annual cycle of ice with that based on the TES data. Figure~\ref{fig:TES_MAOAM_h2o_4_op} (left column) presents the seasonal variations of zonally averaged water vapor and ice from the TES data (MY24--26). The simulated quantities are shown on the right panels. It is seen that the water vapor retrieved from the TES observations is close to that measured by SPICAM \citep{korablev2006water}. The bottom panels present the column opacity of water ice clouds at 12 micron. To calculate the cloud opacity $\tau_\textrm{abs}$ of ice particles from the MGCM output, we use the equation from the works of \citet{montmessin2004origin} and \citet{warren2006visible}:
	\begin{linenomath*}
		\begin{equation}
		\label{eq:tau_abs}
		\tau_\textrm{abs} = \frac{3 Q_\textrm{abs}(r,\lambda) M_c}{4 \rho_\textrm{ice} r},
		\end{equation}
	\end{linenomath*}
	where the absorption efficiency $Q_\textrm{abs}(r,\lambda)$ depends on ice particle sizes $r$ and the wavelength $\lambda$, $M_c$ is the integrated cloud mass predicted by the model in kg m$^{-2}$ and $\rho_\textrm{ice}$ is the ice density. The model opacity is compared with the quantities obtained from the TES retrievals \citep{smith2002annual}. It is seen that the model demonstrates a good agreement with the observations. The simulations show no ice clouds over the north polar cap during the aphelion season and predict a dense north polar hood during the perihelion season, which could not be measured by TES. The model reproduces the aphelion equatorial cloud belt, yet slightly displaced northwards compared to observations.
	
	
	\subsection{Longitudinal Variations}
	
	The seasonal and spatial coverage of SPICAM data allows for comparing both zonally averaged water cycle and longitude-latitude maps of water vapor column density with MGCM results. The examples of such maps, where SPICAM data are averaged over 5 Martian years, are presented in Figure~\ref{fig:compare_h2o_lon-lat_interp}. Leaving aside quantitative differences mentioned in Section~\ref{sec:annual}, one may notice the similar variations in longitudes between $L_s = 90^{\circ}$ and 150$^{\circ}$. These structures are present, although less apparent, at other seasons as well. For instance, on the first panel (averaged over the season $L_s = 0^{\circ}$ -- 30$^{\circ}$), the structures are located near the equator between 90$^{\circ}$W and 140$^{\circ}$W (see also Figure~\ref{fig:h2o_lon_lat_globe_ls120}) and northward between 0$^{\circ}$E and 50$^{\circ}$E. The structures are clearly visible in Figure~\ref{fig:compare_h2o_lon-lat_interp}d. It is evident that their extremes occur between 40$^{\circ}$N and 70$^{\circ}$N. These peaks are very robust and stable for more than one hundred Martian sols of simulations.
	
	\begin{figure}[h]
		\setcounter{figure}{6}
		\includegraphics[width=30pc]{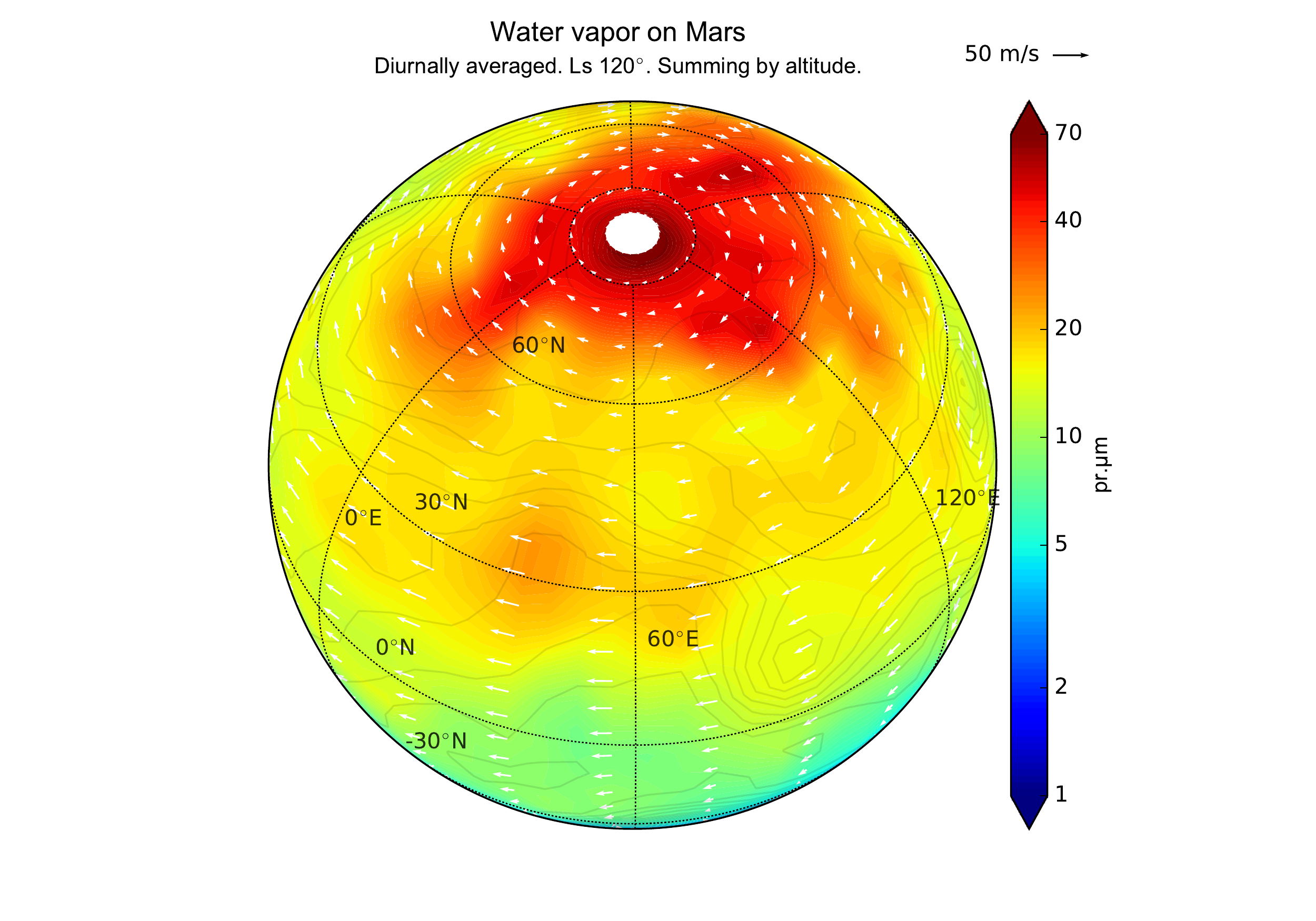}
		\caption{Diurnally averaged water vapor column density in precipitable microns (log-scale) in orthographic projection over the season $L_s \sim 120^{\circ}$ (northern hemisphere summer). Arrows show average horizontal wind velocities in m s$^{-1}$.}
		\label{fig:h2o_lon_lat_globe_ls120}
	\end{figure}
	
	These quasi-symmetric (with respect to the north pole) peaks are most likely associated with stationary planetary waves \citep{richardson1999general, banfield2003forced, haberle2017atmosphere}. Planetary waves in the atmosphere are caused by inhomogeneity of the topography and surface characteristics (albedo and thermal inertia), whose longitudinal structure on Mars is dominated by zonal wavenumbers 2 and 3. The same longitudinal harmonics are clearly seen in Figure~\ref{fig:compare_h2o_lon-lat_interp}. The numerical experiments show that the peaks start to form near the season $L_s\sim 90^\circ$ and disappear later (around $L_s\sim 150^{\circ}$). The main peak is located near 140$^{\circ}$W, in a good accordance with the observations. It should be recalled that the major highland (Olympus Mons and Tharsis Montes) is located close to these longitudes. The other two wave peaks are less stable, and migrate between 0$^{\circ}$ and 130$^{\circ}$E.
	
	
	\subsection{Vertical Distribution}
	\label{sec:vertical}
	
	In the solar occultation mode, SPICAM IR (1.38 $\mathrm{\mu}$m-band) channel has conducted measurements of 82 vertical water profiles during MY28 \citep{betsis2017aerosol, fedorova2018water}. The eclipses were recorded from $L_s \sim 255^{\circ}$ to $L_s\sim 300^{\circ}$, 49 of them in the northern hemisphere and 33 in the southern one. According to the solar occultation technique, each profile corresponds to specific time and location. We have chosen the interval between $L_s = 255^{\circ}$ and $L_s = 267^{\circ}$ for comparison to exclude the period with the major dust storm of MY28. The results for 6 profiles from that period are shown in Figure~\ref{fig:ProfilesComparison} (see also Table~\ref{table:spicam_orbits}).
	
	Figures \ref{fig:ProfilesComparison}a and \ref{fig:ProfilesComparison}d present a comparison of the observed and simulated water vapor density. The model output was averaged over one hour around the corresponding local times (given in Table~\ref{table:spicam_orbits}). Blue lines are for the same as SPICAM data longitudes, while yellow lines show the model results longitudinally averaged in addition. Because of the limited amount of available measurements, we cannot verify whether the variations predicted by the model near the surface are realistic. Above 60--70 km, the uncertainty of retrievals becomes too large. In the middle atmosphere between 40--60 km, the simulated profiles are in a good agreement with the SPICAM retrievals. We admit that the selection of a few SPICAM profiles may be not representative, but a more comprehensive comparison can be a subject of further studies. Figures \ref{fig:ProfilesComparison}b and \ref{fig:ProfilesComparison}e present the simulated water ice particle number density for the sum over all ice particle bins. It can be seen that the ice particle number density is systematically underestimated by the model. As the result, the simulated particle radii are too large compared to observations for some orbits and altitudes (Figures~\ref{fig:ProfilesComparison}c and \ref{fig:ProfilesComparison}f). The reason for this discrepancy is not fully understood. It can be because of the insufficiently accurate parametrization of the bimodal distribution in this season, which lead to the insufficient number of CCN in the model, the slow nucleation rate, or too rapid condensation due to the parameters chosen in the equations (\ref{eq:J_het})--(\ref{eq:drdt}) or the microphysical time step (will be explained in the section \ref{sec:mono}). Despite of that, the comparison is generally favorable, especially for vapor where the Pearson correlation coefficient between the observational and model data does not fall below 0.91. Particularly interesting is that the model tends to reproduce the shape of the profiles of water vapor and ice for particular local times. Disagreements in particle sizes occur mostly for cells with a small amount of water and have very little effect on the whole cycle.
	
	Finally, to complete our comparison with observations, we focus on the ice mixing ratio retrievals. For that, we use 410 assembled limb-viewing observations from CRISM \citep{guzewich2014vertical} to verify the compliance of the MGCM simulations with the experiment. To retrieve the CRISM total column optical depth $\tau(\lambda)$, the following equation was used by \citet{smith2013vertical}:
	\begin{linenomath*}
		\begin{equation}
		\label{eq:norm_opacity}
		\tau(\lambda) = \sum\limits_{i=1}^\textrm{layers} n_i Q_\textrm{ext}(\lambda) \Delta p_i / p_\textrm{surf},
		\end{equation}
	\end{linenomath*}
	where the sum was taken over all atmospheric levels, $n_i$ is the ice mixing ratio, $Q_\textrm{ext}$ is the extinction coefficient, $\Delta p_i$ is the difference in atmospheric pressure across the model layer, and $p_\textrm{surf}$ is the surface pressure. Thus, the retrieved mixing ratios are normalized to the pressure. Combining (\ref{eq:norm_opacity}) and (\ref{eq:tau_abs}), we obtain the method for calculating mixing ratios from the model output.
	
	Figure~\ref{fig:CRISM_mix_full} compares the ice mixing ratios normalized to pressure from simulations and CRISM observations. The data are averaged over longitude and season $L_s\sim 40^{\circ} - 110^{\circ}$. The top and middle panels show the results from CRISM and simulations, respectively. It is seen that the major part of water ice is located above the equator at $\sim$ 20--40 km. This so-called aphelion cloud belt (ACB) regularly occurs over the mentioned season. During the northern spring, the north polar cap starts to sublime. Then, winds transport the vapor to the equatorial regions, where air is still not too warm. This leads to nucleation on the existing dust particles, and formation of the clouds. It is seen that the MGCM reproduces this cycle well despite a few deficiencies, which we focus at below.
	
	Firstly, the model produces slightly thinner ice clouds, especially in the center of the belt. Secondly, the model clouds are $\sim$5 km lower than the CRISM data show. Possible reasons for these discrepancies could be smaller simulated vertical velocities and some excess of large ice particles, as was mentioned in the previous sections. The vertical transport in low latitudes is weak at this season, whereas the large size of ice particles leads to an enhanced sedimentation (and cloud lowering) as well as to a reduced transparency. Nevertheless, the shape and location of the clouds agree well with the observations. It is also seen that most of the ice is spread between 45$^{\circ}$S and 60$^{\circ}$N, which is supported by the observations. The belt has a halo in the center and two thinner tails. In addition, it has some slope in the latitude-altitude plane: ice particles are located higher near the north pole, then the cloud layer altitude drops towards the equator and rises again closer to the south pole. This smile-like shape, which results from a combination of temperature distribution and sedimentation, is successfully reproduced by the model. Some northward offset of the halo center can be caused by the location of the main source of sublimation in the season. The model reproduces lower atmospheric polar clouds above south pole, which were not observed by CRISM.
	
	
	\section{Comparison With the Mono-modal Distributions}
	\label{sec:mono}
	
	Having compared the simulations versus the available observations and verified the ability of the model to reproduce the hydrological cycle, we turn to a comparison of different model scenarios: three with the mono-modal and one with the bimodal dust particle size distributions. Under the first mono-modal scenario M1, the number of large particles is the same as in the bimodal one, but there are no small particles at all. Thus, the total number of all particles in M1 is less than in the bimodal case. The rationale for this scenario is based on the fact that the available observations constrain the number of large particles, while the number of small particles is less known, and one can hypothesize a total lack of them. The second mono-modal scenario M2 includes the same total number of particles as the bimodal one, but also without small particles. To obtain it from the bimodal case, we simply set $\gamma = 0$. We do not consider the mono-modal scenario with only small particles, because it has no observational justification. The third mono-modal scenario M3 is the same as M1, but the microphysical code was turned on only every tenth model time step (every 200 seconds), that is, with a tenfold increase of the microphysical time step. We postpone the consideration of this scenario until later. The simulations under these mono-modal scenarios have been performed similarly to the bimodal one described in the previous section.
	
	The comparison of the annual water cycle was done in a similar manner as in Figure~\ref{fig:TES_MAOAM_h2o_4_op}. For that, we plotted the seasonal dependence of the zonally averaged water vapor and ice in Figure~\ref{fig:maoam_bimodal_h2o_4_op}. It is seen that the water vapor distributions are similar for all scenarios. This occurs because the vapor is strongly controlled by temperature. Therefore, changing the modality of dust scenario has only a slight integral effect on the vapor. However, the distribution in Figure~\ref{fig:maoam_bimodal_h2o_4_op}e corresponding to the M2 scenario looks more diffused than the bimodal one (Figure~\ref{fig:maoam_bimodal_h2o_4_op}a). The distribution for the M3 mono-modal scenario (Figure~\ref{fig:maoam_bimodal_h2o_4_op}g) is very similar to that for the M2 run, and the M1 mono-modal scenario produced somewhat smaller amount of vapor (Figure~\ref{fig:maoam_bimodal_h2o_4_op}a). These discrepancies can be explained by water mass exchanges between ice and vapor. The larger amount of dust particles facilitates the exchange, because it leads to more ice particles. This can explain the difference between the M1 and bimodal cases. However, the simulations also reveal a significant effect of the microphysical time step. As it was mentioned above, M1 and M3 runs are  absolutely identical except that in the M3 case, the microphysics was ran only every tenth dynamical time step. We next consider the effect of the microphysical time step on the whole water cycle.
	
	\begin{figure}[h]
		\setcounter{figure}{10}
		\includegraphics[width=35pc]{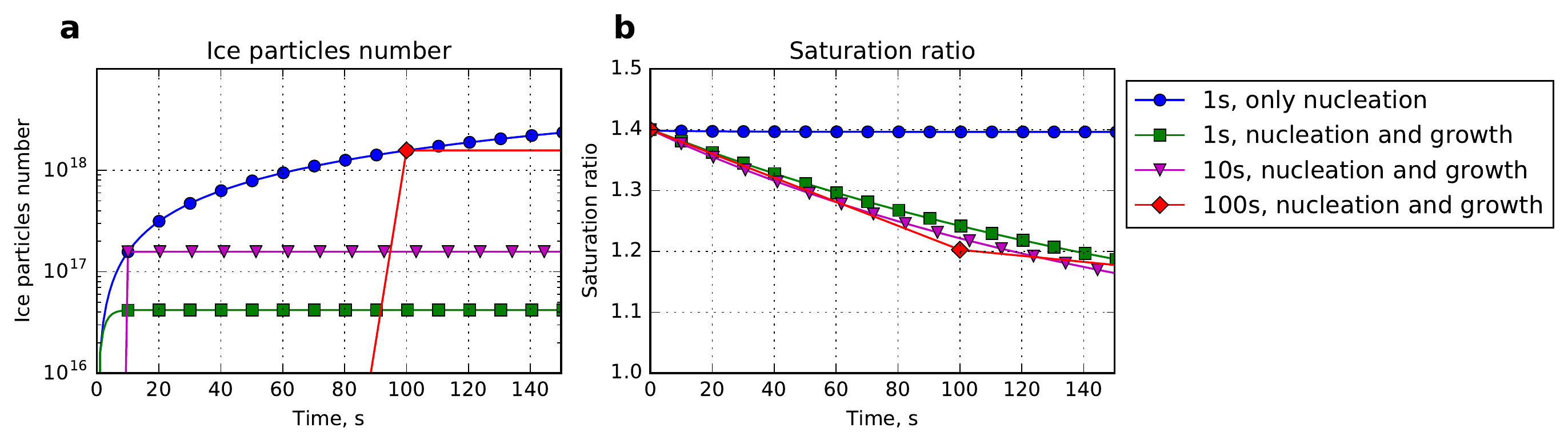}
		\caption{Evolution of (a) the number of water ice particles for the CCN size bin of 0.1 $\mathrm{\mu}$m and (b) water vapor saturation ratio in an explicit time-stepping microphysical scheme in an isolated grid cell (equator, altitude $\sim$ 30 km, $L_s$ $\sim$ 100$^{\circ}$). Advection, diffusion and any water mass exchanges with other cells have been disabled. The initial ice mass and number of \myadded{ice} particles \mydeleted{for all CCN }were set to zero. Solid color lines denote the results for different microphysical time steps from 1 to 200 seconds (see the legend). The blue line corresponds to the results with disabled ice particles growth (only nucleation was permitted). In the other experiments, ice particles can nucleate and grow.}
		\label{fig:maoam_micropys_steps}
	\end{figure}
	
	In order to illustrate processes occurring when the microphysical time step is long, we consider a grid cell with disabled water mass exchanges with the surrounding cells, i.e., with disabled advection, sedimentation and diffusion. The initial amount of ice mass and number of ice particles were set to zero and the results of calculations are presented in Figure~\ref{fig:maoam_micropys_steps}. The resulting water supersaturation led to fast nucleation and particle growth (purple line). We also ran the code without particle growth, only with nucleation (blue line). It is seen that the microphysical time steps longer than one second yield an instantaneous nucleation. The particle growth on the second step produces a sharp reduction of the water vapor mass, decrease in the water vapor saturation ratio (Figure~\ref{fig:maoam_micropys_steps}b) and, eventually, leads to a complete cessation of nucleation. Using smaller microphysical time steps can slow down this process, thus reproducing it more realistically. The formation of a smaller number of ice particles decelerates the decrease of the water vapor mass, which, in turn, slows down the nucleation. With time, the number of ice particles tends to a constant value, because the nucleation stops and mass exchange with other grid cells was disabled. Therefore, longer microphysical time steps tend to increase the amount of simulated ice particles (Figure~\ref{fig:maoam_micropys_steps}a). During a single step, the amount of formed ice particles becomes as large as if the nucleation was not slowed down by the particle growth during this time. This happens because we calculate the nucleation and particle growth consequently. The same approach is used in other MGCMs \citep[e.g.][]{navarro2014global} and apparently produces the same effect (i.e., dependence of the results on the microphysical time step).
	
	The right column of Figure~\ref{fig:maoam_bimodal_h2o_4_op} presents the comparison of the water ice cloud opacity in the atmospheric column during one Martian year. All the shown values are diurnally and zonally averaged. The M1 mono-modal dust scenario produces very thin ice clouds (Figure~\ref{fig:maoam_bimodal_h2o_4_op}d), which are caused by an insufficient amount of CCN in the atmosphere. Figures~\ref{fig:maoam_bimodal_h2o_4_op}d and \ref{fig:maoam_bimodal_h2o_4_op}f show that a larger number of CCN leads to a greater amount of ice mass in the atmosphere, thus increasing the column opacity of ice. The amount of large CCN in the M2 scenario is greater than it was observed in the Martian atmosphere, and the simulated cloud opacities are two times greater than in the measurements (Figure~\ref{fig:maoam_bimodal_h2o_4_op}b). It is seen that the tenfold increase of the microphysical time step in M3 also approximately doubled the simulated column ice opacity, although the latter is still below the observed quantities (Figure~\ref{fig:TES_MAOAM_h2o_4_op}c). The M3 simulation illustrates that application of longer time steps can bring model results closer to observations, however this process \added{is} caused by purely numerical aspects and has no physical ground. On the contrary, the bimodal dust scenario produces better agreement with the measurements (Figure~\ref{fig:maoam_bimodal_h2o_4_op}b), although the utilized microphysical time step (20 seconds) should probably be decreased in future applications. The distributions of opaque clouds over winter poles are similar in the simulations despite the mentioned differences in the magnitudes. Also interesting is that the greater number of large particles (M2 scenario) produces the aphelion cloud belt not symmetric with respect to the seasons.
	
	Figure~\ref{fig:maoam_bimodal_profiles} presents the vertical profiles of (averaged over a Martian year) water ice characteristics simulated using the M1 mono-modal and bimodal dust particle size scenarios. Note that identical CCN sizes are marked by the same colors. As it was described in section \ref{sec:bimod-dust}, the ice mass and number density increase in every CCN bin separately (the so-called two-moment scheme). One can see that the amount of ice mass in the M1 mono-modal dust scenario (Figure~\ref{fig:maoam_bimodal_profiles}b) is much smaller than in the bimodal one (Figure~\ref{fig:maoam_bimodal_profiles}a) for all bins except in the bin with the radius 0.3 $\mathrm{\mu}$m. This bin becomes the main source of ice mass in the M1 mono-modal scenario. This is caused by the choice of parameters of the dust distribution. For the same reason CCN with an average size 0.03 $\mathrm{\mu}$m become the main source of ice mass for the bimodal scenario. This is not surprising because these CCN correspond to the first peak of the bimodal distribution (see, e.g., Figure~\ref{fig:maoam_bimodal_evolution}). The amount of ice mass in the bimodal scenario is greater than in the M1 case, because of the number of ice nuclei. As was discussed above, the number of small dust particles is approximately 10$^3$--10$^4$ times greater than the number of large particles. Such abundance of condensation nuclei allows for catching much more vapor in the bimodal simulation and, as a result, to produce heavier clouds. In addition, the larger number of ice particles results in a decrease of their mean radius, which reduces sedimentation and leaves more ice airborne.
	
	It is easy to see from Figures~\ref{fig:maoam_bimodal_profiles}c and \ref{fig:maoam_bimodal_profiles}d that the number density of ice particles corresponding to the small CCN in the M1 mono-modal experiment is approximately an order of magnitude smaller than in the bimodal.  The effect of the bimodal distribution is best illustrated by the last two panels, where the ice particle effective radii averaged over a Martian year are shown for the both simulations. Because of the small amount of dust particles covered by ice in the M1 mono-modal scenario, the resulted average radii reach 10--15 $\mathrm{\mu}$m in the middle atmosphere, which absolutely disagrees with the observations \citep{fedorova2014evidence, guzewich2014vertical}. Otherwise, the bimodal scenario produces much smaller ice particles for the CCN with small cores in a good agreement with the measurements, as it was discussed in section~\ref{sec:vertical}. Some local excesses of the radii are compensated by averaging. It should be mentioned that, even in the bimodal scenario, we obtain some surplus of ice radius sizes, especially near the surface, where it could be caused by the insufficiently rigorous sublimation/condensation scheme. Hence, a further development of the model should address this limitation. Based on the foregoing, we can state with certainty that the use of the bimodal distribution with an added peak of small particles increases the number of nucleation particles in the atmosphere contributing to the growth of the ice mass, increases the concentration of ice particles and reduces their radii, which, in turn, improves the simulated opacity of the clouds compared to observations.
	
	Finally, after considering the effects of the bi-modality of the dust distributions on the annual cycle, we turn our attention to the details of the cycle. For that, we consider again the CRISM observations. They are presented in Figure~\ref{fig:CRISM_mix_full} along with the latitude-altitude distributions of ice clouds simulated with bi- and M1 mono-modal scenarios. First, it is seen that there are less clouds over the south pole in the mono-modal scenario. It is an interesting result, which points at the insufficient amount of CCN in the M1 mono-modal scenario above the south pole during the seasons $L_s\sim 40^{\circ} - 110^{\circ}$. Potentially, some clouds nucleated on small particles can occur there, which can be proved or refuted by future Mars missions. Second, note that the lower panel is plotted with a different scale. The displayed ice mixing ratio for the aphelion cloud belt in the M1 mono-modal scenario is approximately three times smaller than that in the bimodal one. The reasons for that are described in the previous paragraph. Additionally, huge sizes of the particles lead to a strong sedimentation, which confines the clouds at levels closer to the surface. This effect is illustrated by Figure~\ref{fig:CRISM_mix_full}c showing that the clouds are located lower than in Figure~\ref{fig:CRISM_mix_full}b. Generally, the shape of the clouds simulated with the mono-modal scenario, their sizes and the specific meridional slope of the cloud belt are close to those in the bimodal experiment.
	
	
	\section{Summary and Conclusions}
	\label{sec:conclusion}
	
	We presented a new implementation of the hydrological cycle scheme in the Martian atmosphere into the Max Planck Institute Martian general circulation model (MPI-MGCM, also known as MAOAM). The hydrological model includes a semi-Lagrangian scheme for describing transport of water vapor and ice particles, and a microphysical model for calculating phase transitions between them. The microphysical scheme generally follows the approach of \citet{montmessin2002new, montmessin2004origin} and \citet{navarro2014global}, but utilizes different microphysical parameterizations, which are described in section~\ref{sec:MGCM}. Simulations with MPI-MGCM have been performed using a predetermined distribution of dust that represented a seasonal and spatial evolution of atmospheric aerosol based on the Thermal Emission Spectrometer onboard Mars Global Surveyor (MGS-TES) and the Planetary Fourier Spectrometer onboard Mars Express (MEX-PFS) measurements with the global dust storms removed. The number of particles (serving as nuclei) of particular size (subdivided into bins) was calculated using the bimodal log-normal dust distribution (the distribution is called bimodal if its density function has two peaks). Unlike in the previous works \citep[e.g.,][]{montmessin2002new}, we employed a distribution that included much smaller particles and the ratio of number density peaks up to 10$^3$--10$^4$, which are apparently supported by observations for certain seasons \citep{fedorova2014evidence}. The hydrological scheme also includes saturation, nucleation, particle growth, sublimation and sedimentation depending on the average radius of particles and the surface microphysics.
	
	The simulated annual variations, horizontal and vertical distributions of water vapor and ice clouds have been compared with SPICAM (the Spectroscopy for Investigation of Characteristics of the Atmosphere of Mars onboard Mars Express), MGS-TES and CRISM (the Compact Reconnaissance Imaging Spectrometer for Mars onboard Mars Reconnaissance Orbiter) observations. The simulated general amount of vapor in the summer season at the north pole with the maximum near 50--70 precipitable microns is close to the SPICAM observations. The simulated vapor is absent in cold seasons during north and south winters, thus indicating that the conversion between vapor and ice as well as their transport are captured realistically. We showed that the model is able to replicate the horizontal distribution of water vapor with the quasi-symmetric with respect to the north pole peaks of concentration between $L_s = 90^{\circ}$ and 150$^{\circ}$. These structures are most likely associated with stationary planetary waves \citep{richardson1999general, banfield2003forced, haberle2017atmosphere}.
	
	The comparison of the model results with the SPICAM profiles demonstrated a good agreement for water vapor in the middle atmosphere (40--60 km), and a systematic underestimation of the simulated ice density. However, the latter disagreement occurred mostly for cells with a small amount of water and, thus, affected the whole cycle very little. The model also predicted an overly dense north polar hood during the perihelion season, which largely could not have been measured by TES, and prevented cloud formation right above the north polar cap during the aphelion season, consistently with the observations. The simulated aphelion cloud belt (water ice clouds above the equator during the seasons $L_s\sim 40^{\circ} - 110^{\circ}$) is in a good agreement with the CRISM observations.
	
	Simulations for mono-modal and bimodal particle size distributions demonstrated that the latter scenario most strongly affects the modeled ice clouds mass, opacity, number density and particle radii bringing them closer to observations. The simulations showed much weaker effect of the bi-modality (excess of small aerosol particles) on water vapor distributions.  The use of the second peak of small CCN in the bimodal distribution increases the number of particles nucleated in the atmosphere, which contributes to the growth of ice mass, increases the concentration of ice particles and reduces their radii. This, in turn, improves the simulated opacity of the clouds compared to observations.
	
	More generally, our results highlight the importance of the dust size distribution with the peak of small particles for modeling water ice in the atmosphere of Mars. \added{We must cautiously state that this result may be a model-dependent and can also be affected by the hypotheses and parameterizations made in our model (e.g, the lack of interactions between dust and water ice).} Also, it is reasonable to expect that these distributions throughout all seasons and locations are not perfectly bimodal, but have more complex shapes. More measurements and MGCM simulations that self-consistently account for dust transport \citep[e.g.,][]{navarro2014global,kahre2017updates} can further clarify this \added{and shed the light on the modality of the dust size distribution}.


	\acknowledgments
	
	The data supporting the MPI-MGCM simulations can be found at \underline{https://\-mars.mipt.ru}, \added{\underline{https://\-zenodo.org/record/1045331}} \citep{shaposhnikov2017modeling} or obtained from D. Shaposhnikov (shaposhnikov@phystech.edu).
	
	We express our deep gratitude to Alexander Trokhimovskiy, Daria Betsis, Chris Mockel and Scott Guzewich for assistance with the observational data and helpful discussions.
	
	This work was performed at the Laboratory of Applied Infrared Spectroscopy of Moscow Institute of Physics and Technology and at Max Planck Institute for Solar System Research. The work was partially supported by the Russian Science Foundation grant 100027.07.32.RSF27 and German Science Foundation (DFG) grant ME2752/3-1.
	

	

	
	\begin{table}[h]
		\caption{Characteristics of the SPICAM orbits}
		\centering
		\begin{tabular}{*{5}{c}}Orbit & Longitude (E) & Latitude (N) & Local time (h) & $L_s$ \\
			\hline4407A3 & 54.19 & -25.75 & 5.14 & 254.94 \\4409A2 & 215.82 & -28.63 & 5.03 & 255.30 \\4421A1 & 168.45 & 41.52 & 7.55 & 257.41 \\4428A1 & 207.70 & 46.97 & 7.92 & 258.65 \\4435A1 & 247.16 & 51.49 & 8.31 & 259.89 \\4461A2 & 109.05 & -60.03 & 2.34 & 264.53 \\
			\hline
		\end{tabular}
		\label{table:spicam_orbits}
	\end{table}
	
	\begin{figure}[h]
		\setcounter{figure}{2}
		\centerline{\includegraphics[width=40pc]{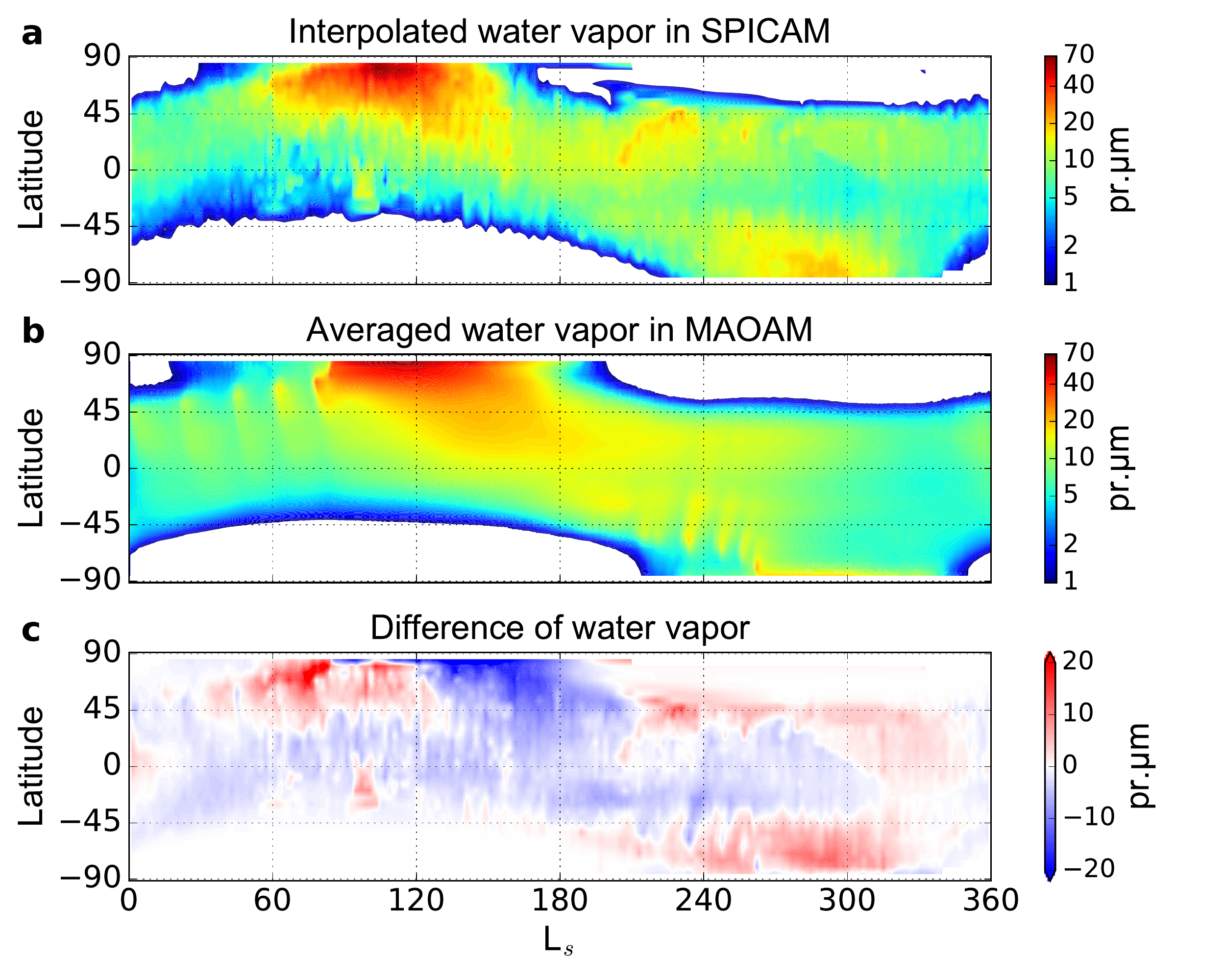}}
		\caption{Annual water cycle on Mars from the SPICAM data \citep{trokhimovskiy2015mars} and simulated with the MPI-MGCM. The top panels show water vapor column density a) observed with SPICAM averaged over 5 Martian years and zonally, and b) simulated with the MGCM averaged daily and zonally. The lower panel presents the difference between the SPICAM data and the simulations.}
		\label{fig:compare_h2o_sum}
	\end{figure}
	
	\begin{figure}[h]
		\setcounter{figure}{3}
		\centerline{\includegraphics[width=40pc]{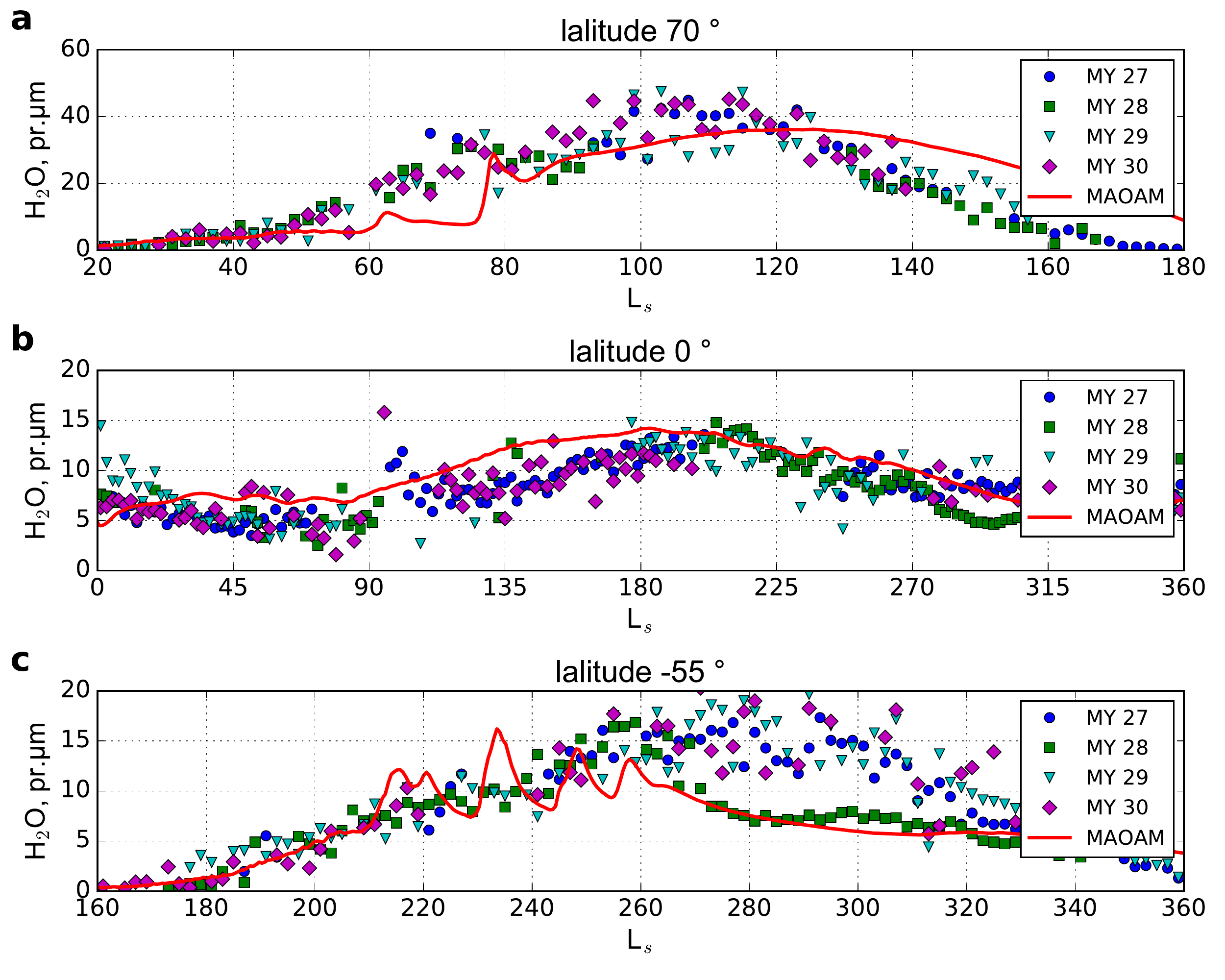}}
		\caption{Annual water cycle measured by SPICAM (markers) \citep{trokhimovskiy2015mars} and simulated with the MGCM (red lines, averaged daily and zonally) at three characteristic latitudes (70$^{\circ}$N, 0$^{\circ}$ and 55$^{\circ}$S). Color and shape of markers indicate the Martian year of observations (see the annotations) } 
		\label{fig:maoam_h2o_lat_compar}
	\end{figure}
	
	\begin{figure}[h]
		\setcounter{figure}{4}
		\centerline{\includegraphics[width=40pc]{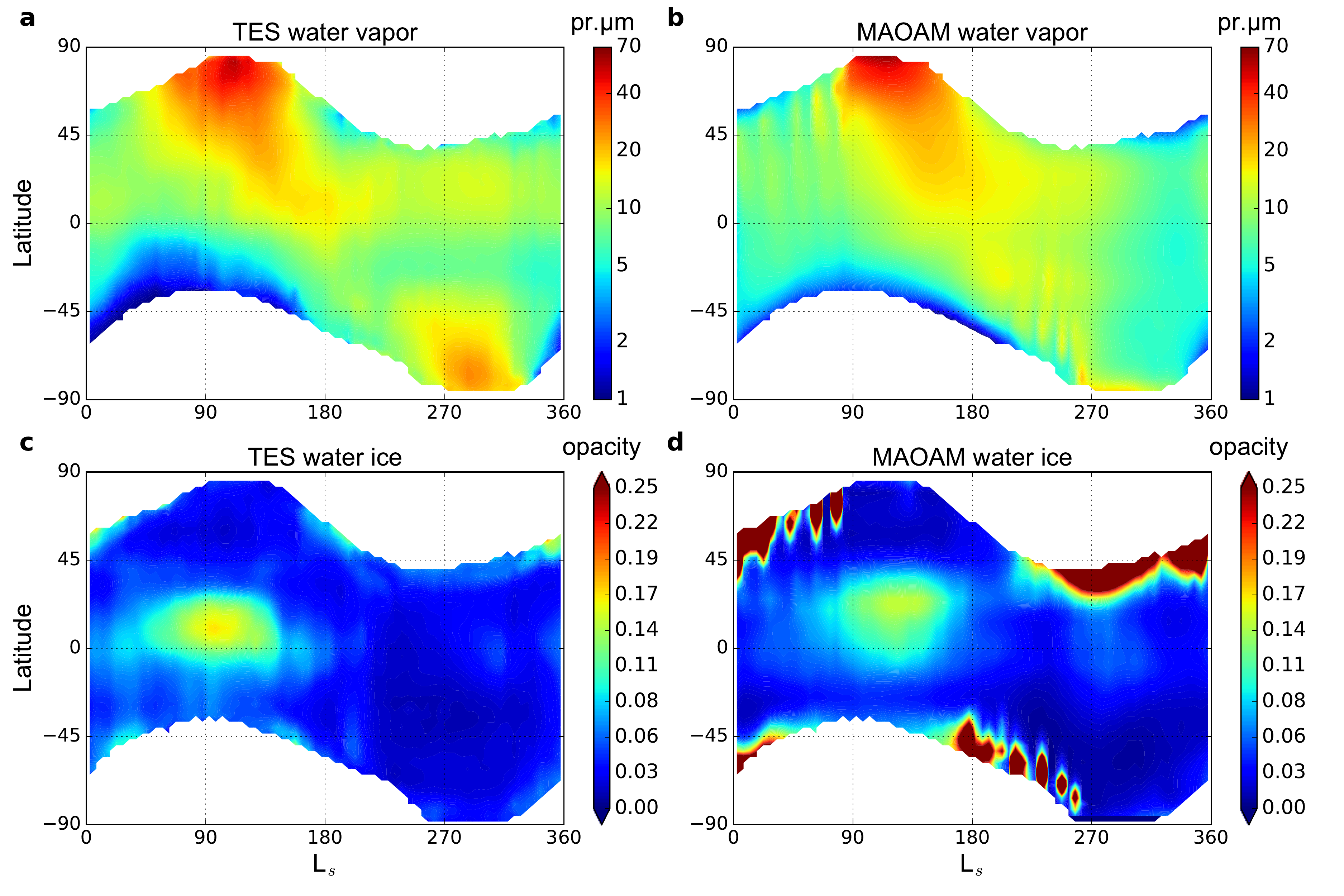}}
		\caption{Comparison of the simulated annual water cycle averaged daily and zonally with TES observations \citep{smith2001thermal} averaged zonally and over 3 Martian years (MY24--26). The upper panels are for the precipitable water vapor from TES (left) and from the model (right). The bottom panels are the same, but for the column opacity of water ice particles at the wavelength 12 $\mu$m.}
		\label{fig:TES_MAOAM_h2o_4_op}
	\end{figure}
	
	\begin{figure}[h]
		\setcounter{figure}{5}
		\centerline{\includegraphics[width=40pc]{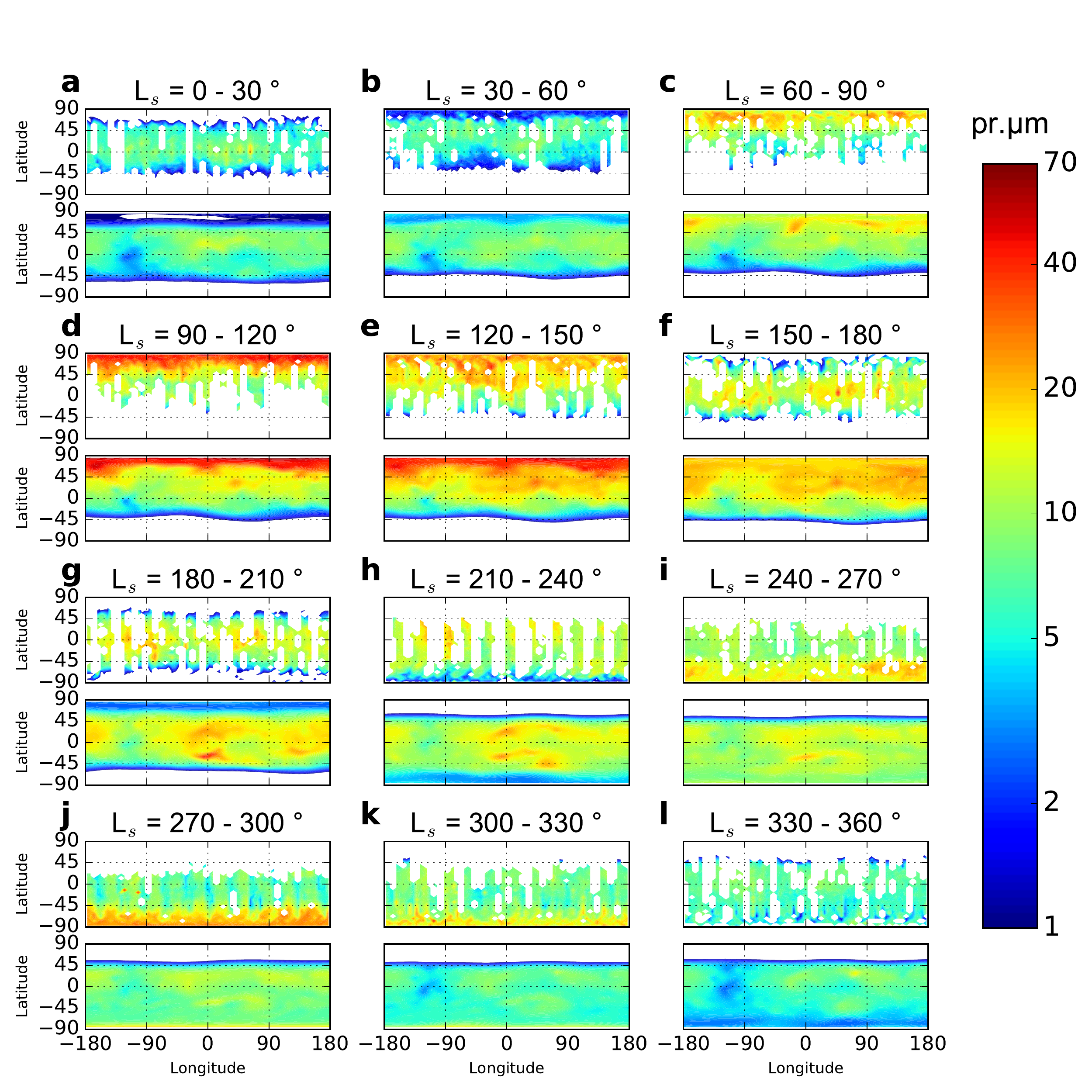}}
		\caption{Horizontal distributions of water vapor column density (in preciptable microns) observed with SPICAM (upper parts of each panel) and simulated with the MGCM (daily averaged). In each pair of plots, the water vapor measured by SPICAM is averaged over 5 Martian years.}
		\label{fig:compare_h2o_lon-lat_interp}
	\end{figure}
	
	\begin{figure}[h]
		\setcounter{figure}{7}
		\centerline{\includegraphics[width=40pc]{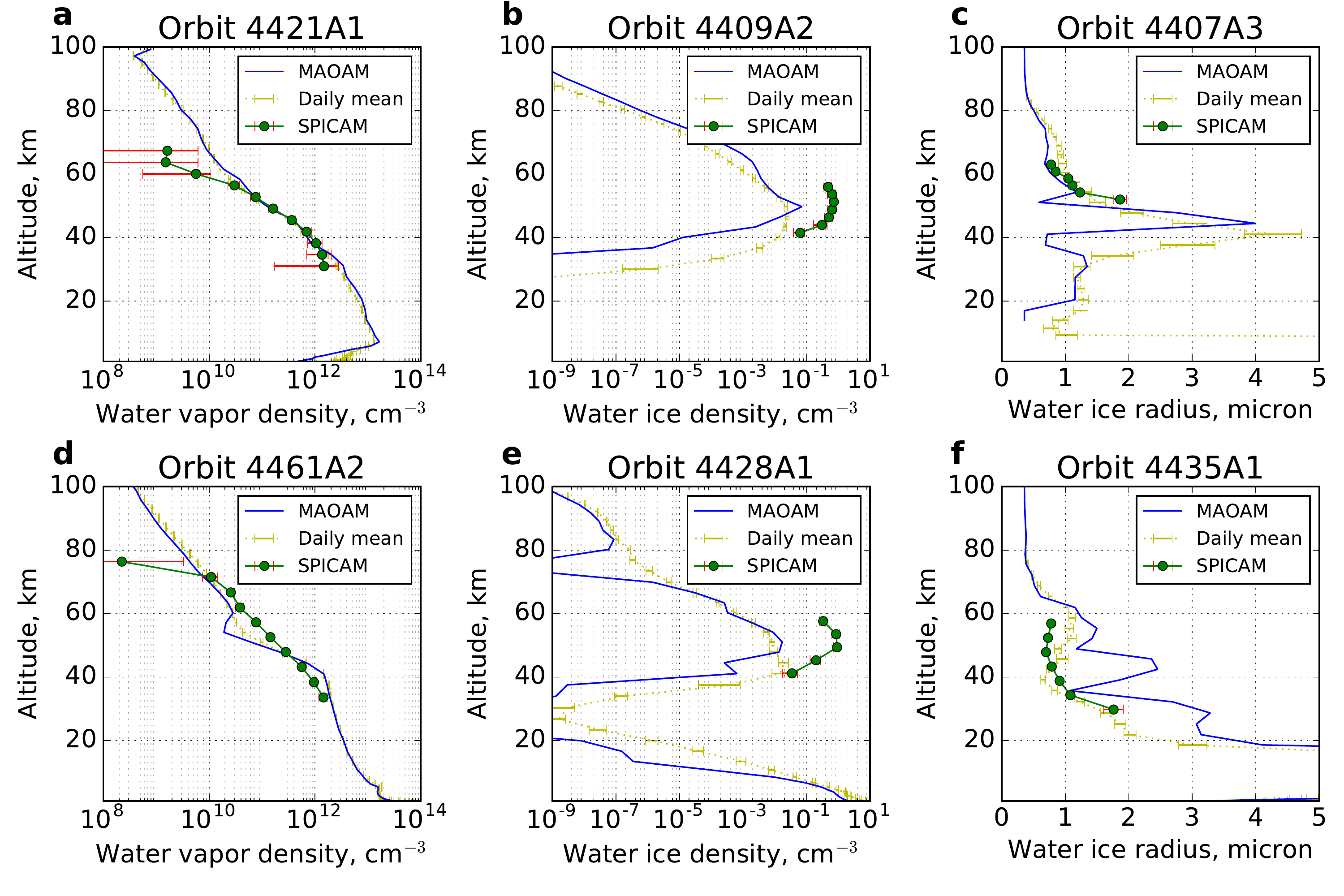}}
		\caption{Vertical profiles of water vapor number density in cm$^{-3}$ (a, d), water ice density in cm$^{-3}$ (b, e) and effective ice radii in $\mathrm{\mu}$m (c, f) from the SPICAM data (MY28, green markers) \citep{betsis2017aerosol, fedorova2018water} and MGCM simulations (blue and yellow lines) for several orbits given in Table~\ref{table:spicam_orbits}. Each orbit corresponds to particular time and location (see the Table~\ref{table:spicam_orbits}). Red errorbars denote the standard deviation of the observations, blue lines indicate hourly averaged simulated quantities on the same as the SPICAM data coordinates and time, yellow dots and errorbars show longitudinally averaged simulations at the local time of observations and the standard errors of the mean of the MGCM results, respectively.}
		\label{fig:ProfilesComparison}
	\end{figure}
	
	\begin{figure}[h]
		\setcounter{figure}{8}
		\centerline{\includegraphics[width=40pc]{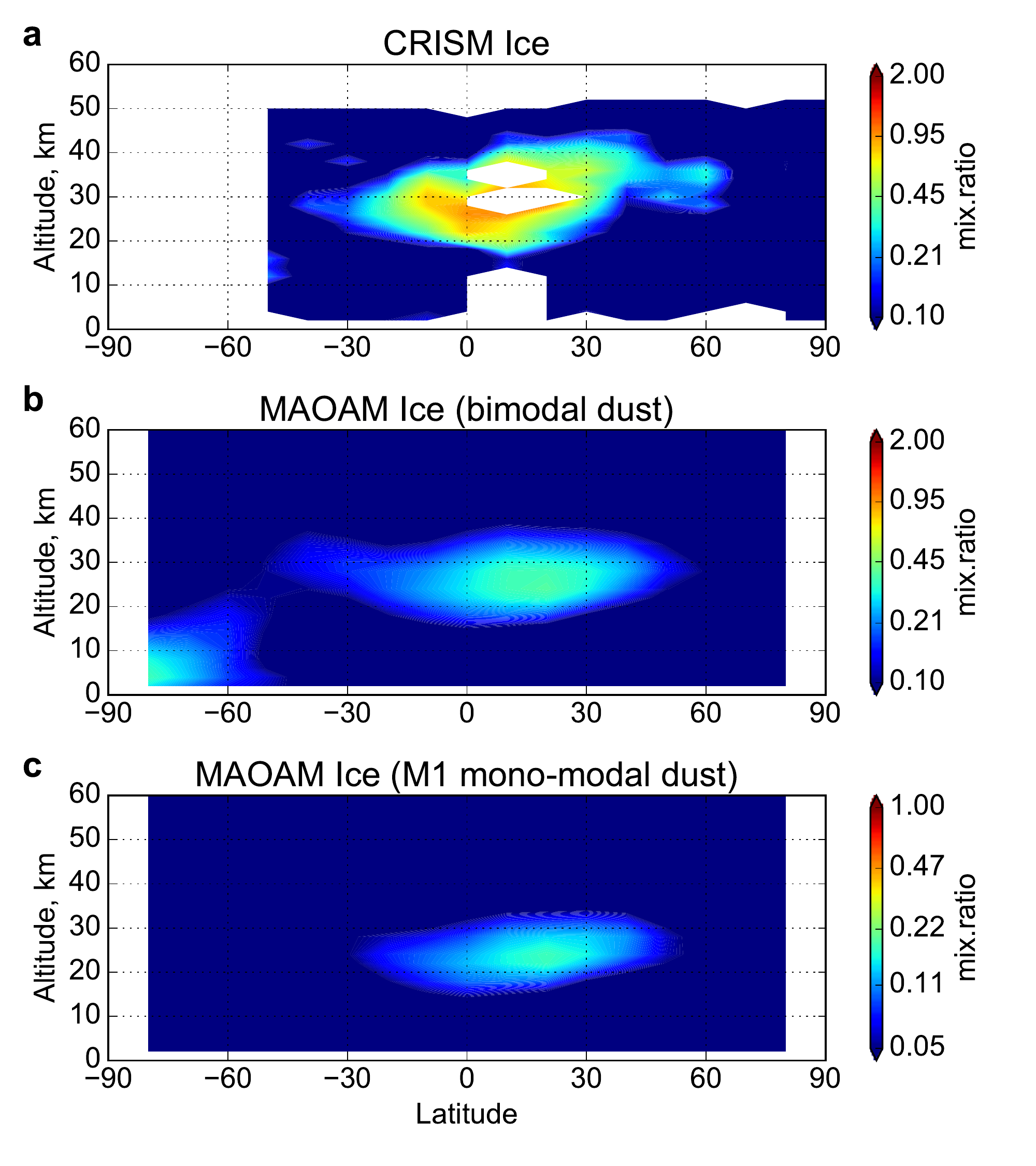}}
		\caption{Comparison of the observed by CRISM (a) \citep{smith2013vertical, guzewich2014vertical} and simulated ice mixing ratios normalized to pressure (see the eq.~(\ref{eq:norm_opacity})) with the bimodal (b) and M1 mono-modal (c) dust size distribution scenarios from the surface up to 60 km averaged zonally and over the season $L_s\sim 40^\circ - 110^{\circ}$. }
		\label{fig:CRISM_mix_full}
	\end{figure}
	
	\begin{figure}[h]
		\setcounter{figure}{9}
		\centerline{\includegraphics[width=40pc]{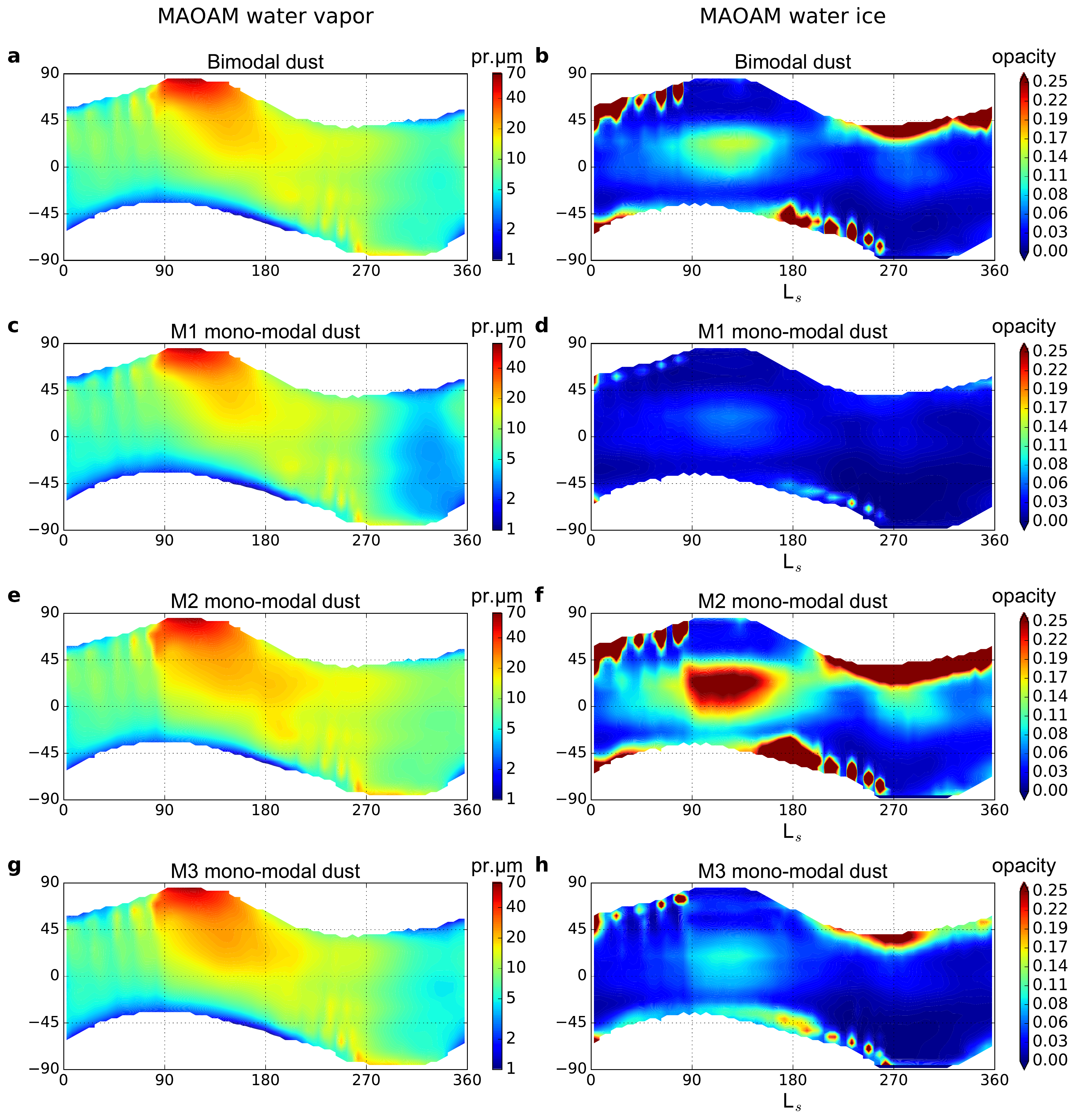}}
		\caption{Comparison of the annual water cycle using the bimodal (a, b), M1 mono-modal (c, d), M2 mono-modal (e, f) and M3 mono-modal (g, h) CCN scenarios (see the section~\ref{sec:mono}). The left panels illustrate the water vapor in precipitable $\mathrm{\mu}$m. The right ones show the column opacity of water ice particles at the wavelength 12 $\mathrm{\mu}$m. The data are daily and zonally averaged.}
		\label{fig:maoam_bimodal_h2o_4_op}
	\end{figure}
	
	\begin{figure}[h]
		\setcounter{figure}{11}
		\centerline{\includegraphics[width=30pc]{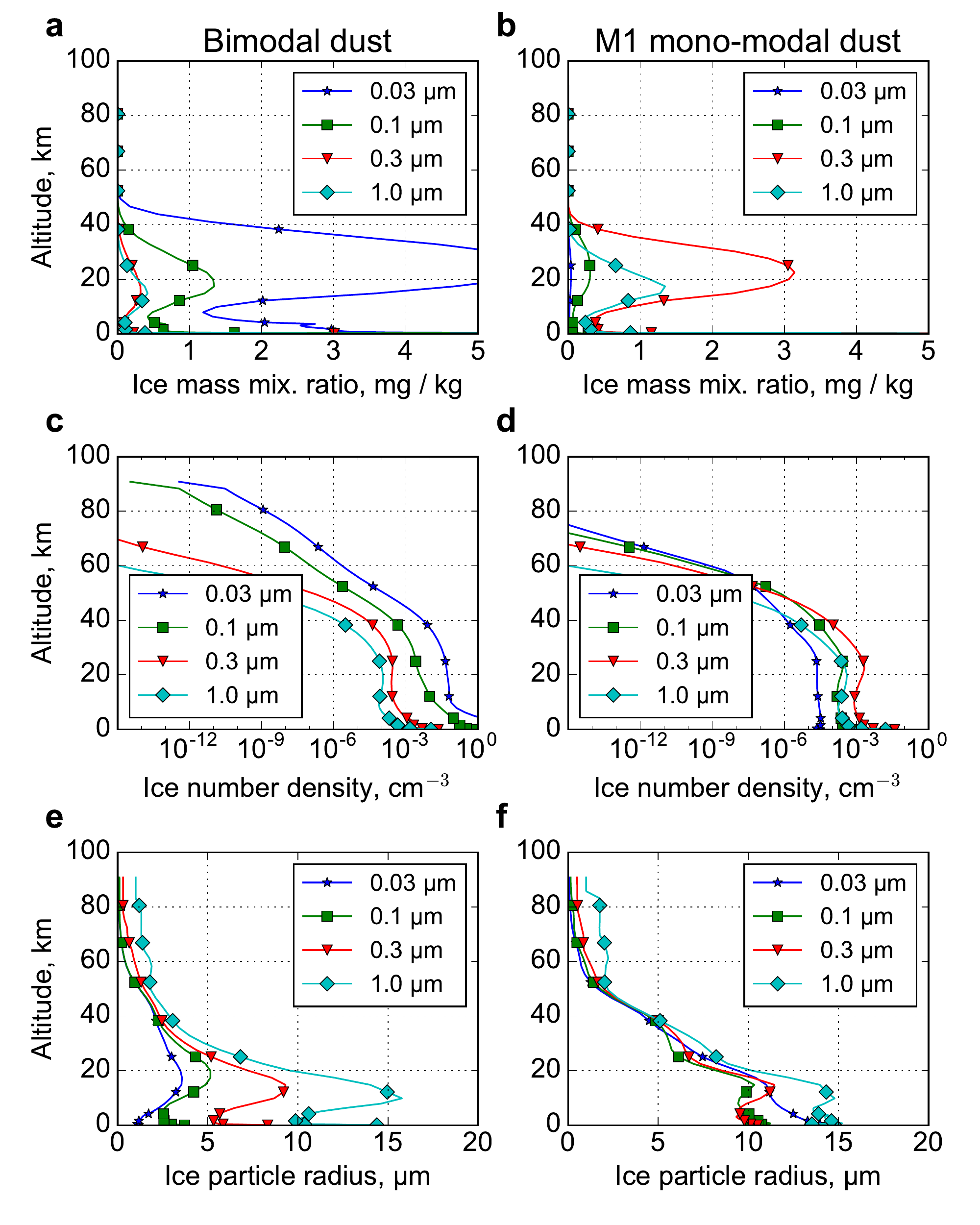}}
		\caption{Vertical profiles of the water ice mass mixing ratio (in mg kg$^{-1}$, upper row), number density (in cm$^{-3}$, central row) and ice particle radii (in $\mathrm{\mu}$m, lower row) simulated with the bimodal (a, c, e) and M1 mono-modal (b, d, f) dust scenarios. All profiles are based on the output averaged over latitude, longitude and seasons. Solid color lines denote the results for 4 different bins: with CCN radii 0.03, 0.1, 0.3 and 1 $\mathrm{\mu}$m (see also the legend for notations).}
		\label{fig:maoam_bimodal_profiles}
	\end{figure}
	
	\listofchanges
	
	\myexplain{Replaced: \mydeleted{characteristics} replaced with: in the model, on page 1.}
		
	\myexplain{Replaced: \mydeleted{[Neary and Daerden, 2017]} replaced with: [Neary and Daerden, 2018], on page 3.}
	
	\myexplain{Replaced: \mydeleted{The initial ice mass and number of particles for all CCN were set to zero} replaced with: The initial ice mass and number of ice particles were set to zero, on page 18.}
	
\end{document}